# Concept-oriented programming:
# from classes to concepts and from inheritance to inclusion

Alexandr Savinov

http://conceptoriented.org/savinov

**Abstract.** For the past several decades, programmers have been modeling things in the world with trees using hierarchies of classes and object-oriented programming (OOP) languages. In this paper, we describe a novel approach to programming, called concept-oriented programming (COP), which generalizes classes and inheritance by introducing concepts and inclusion, respectively.

## 1. Introduction

The basic idea behind object-oriented programming (OOP) is that objects encapsulate state and behavior which are modeled by means of classes. This abstraction works out pretty well because most of the world is hierarchical and in most cases things can be easily fit into a hierarchy of classes. However, the problem is that OOP does not provide any means for describing how objects are *represented* and how they are *accessed*. It is assumed that there is some kind of primitive references which are used for that purpose while the programmer is not able to change them or define new object access methods. References are almost completely removed from the scope of OOP and therefore it is not possible to model how objects exist, where they exist and how they are accessed. This treatment of hierarchies in terms of classes leads to several serious drawbacks:

- Classes allow us to model only objects and their behavior but not references. References have been considered second class citizens in the area of computer programming, data modeling, analysis and design. And the notion of reference behavior has not even been studied, that is, references were supposed to be completely passive elements of the program.

- Classes allow us to model a conceptual hierarchy but not a hierarchy of objects. In other words, class instances (objects) always exist in flat space even though their classes are defined as a hierarchy. Yet, the benefits of having object hierarchies have been demonstrated in prototype-based programming and there is a definite need in supporting object hierarchies in class-based languages.

- Classes do not allow us to model intermediate behavior executed behind the scenes during object access. They also do not support cross-cutting concerns injected in arbitrary points of the program. The advantages and need in these mechanisms were demonstrated in aspect-oriented programming as well as some other approaches like inner methods.

Concept-oriented programming (COP) is a novel programming paradigm the main goal of which is to fill these gaps by *legalizing references* and making them first-class elements of programs. In this sense, COP can be characterized as a reference-oriented approach or programming focusing on what happens during access. References in COP can be defined and manipulated explicitly rather than being hidden in the run-time environment. COP generalizes OOP by revisiting such main notions as class and inheritance, and by assuming that not only objects but also references account for a great deal of the program complexity. In other words, if OOP deals with objects then COP deals with both objects and references.

To model both references and objects, COP introduces a novel programming construct, called *concept* (hence the name of this approach). The main goal of concepts in comparison to classes is to provide a possibility to model how objects are represented and accessed. Concepts generalize classes because under certain simplifying conditions they behave precisely as classes. COP also generalizes inheritance by introducing a new relation, called *inclusion*. The purpose of inclusion relation is to model hierarchical address spaces by describing multi-segment references similar to postal addresses. The main distinction of inclusion from inheritance is that it is able to model containment so that objects exist in a hierarchy rather than in flat space. Essentially, COP assumes that inheritance is a particular case of containment and to be included in some element means to inherit its properties. Concepts and inclusion relation allow us to describe program elements as consisting of two parts (reference and object) and existing in a hierarchical address space. This program structure makes it possible to describe many mechanisms and patterns of thoughts currently belonging to different programming paradigms: modeling object hierarchies in prototype-based programming (Borning, 1986; LaLonde et al., 1986; Lieberman, 1986), precedence of parent methods over child methods in the Beta inner methods (Goldberg et al., 2004), modularizing cross-cutting concerns in aspect-oriented programming (Kiczales et al.,1997), value-orientation in functional programming.

Concept-oriented programming is an emerging technology which has passed through several major revisions starting from the first version, denoted COP-I (Savinov, 2005a), where concept was defined as a couple of one object class and one reference class and inclusion relation was introduced. Object class in COP-I is equivalent to conventional classes. Reference class is aimed at reifying references by explicitly exposing their structure and behavior at the level of the programming language. The next version, COP-II (Savinov, 2007; Savinov, 2008; Savinov, 2009b), changed the role of reference classes and object class in inclusion relation. This paper describes the latest revision of concept-oriented programming, COP-III (Savinov, 2012a). The first major change in COP-III is that concepts are defined differently: instead of using two symmetric constituents – object class and reference class – COP-III uses only one component which models references. In order to compensate for the absence of these two classes, a new mechanism of dual methods is introduced where each method has two implementations, called *incoming* method and *outgoing* method. Another important change is the use of two keywords for navigating through the hierarchy, *super* and *sub* (as opposed to using only super in OOP), and the existence of two opposite overriding strategies.

## 2. Background

### 2.1. Intermediate Behavior and Cross-Cutting Concerns

In OOP, all objects are represented by *primitive* references which provide *direct* access to the object. For example, given a bank account reference, we can call its method `account.getBalance()` for getting the current balance. This method will be started immediately after it has been called without any intermediate actions because it is represented by a primitive reference. COP changes this *instantaneous action* principle by introducing custom references which not only allow for arbitrary representation format but also make it possible to inject any intermediate actions into the mechanism of object access (method calls). In COP, a bank account reference will effectively intercept all requests to the represented object and execute actions according to some custom domain-specific logic of bank account access which is defined by the programmer. For example, we might want to perform security checks, manage transactions or manipulate persistent state of bank accounts. Thus object access (and method calls) is not instantaneous process anymore and a great deal of the program logic is modularized in references rather than in objects.

Importantly, this intermediate behavior cannot be associated with any specific object class because it describes the logic of access rather than the logic of the end methods. In particular, one and the same reference type could be used for representing many different object classes. In this sense, the behavior of references is a typical cross-cutting concern (Kiczales et al.,1997). It is also important to note that intermediate actions associated with references are executed behind the scenes so that the programmer retains the illusion of instantaneous action. The programmer is still able to use objects as if they were directly accessible but at the same time he has the possibility to inject any intermediate code which is triggered automatically for each access. In contrast, OOP does not provide any facilities for describing custom references and intermediate behavior so that all objects are represented and accessed in the same way. COP fills this gap and allows the programmer to effectively separate both concerns (Dijkstra, 1976): explicitly used business logic associated with objects, and intermediate behavior executed implicitly during object access and associated with references.

The problem of indirect object representation and access can be solved by using such approaches as dynamic proxies (Blosser, 2000), mixins (Bracha & Cook, 1990; Smaragdakis & Batory, 1998), metaobject protocol (Kiczales et al.,1991; Kiczales et al.,1993), remoting via some middleware (Monson-Haefel, 2006), smart pointers (Stroustrup, 1991), aspect-oriented programming (Kiczales et al.,1997), variability and feature modeling (Younis, Ghoul & Alomari, 2013; Younis & Ghoul, 2014). What is unique in COP is that it provides a principled language-based solution which generalizes OOP rather than adds a new orthogonal mechanism.

## 2.2. Hierarchical Address Space

In OOP, all objects exist in one flat space where they are represented by one type of primitive references. In COP, a program is viewed as a nested space (Fig. 1). Each internal space is identified with respect to its parent space using a local reference. Thus an element is identified by a sequence of reference segments where each next reference identifies an internal space. Such an identifier is referred to as a *complex references* while its constituents are referred to as *reference segments*. This structure is analogous to the conventional postal addresses. For example, an element in the postal address space could be identified by the following complex address: `<"Germany", "Dresden", "Haupt Str. 25">`. An important consequence of such design is that an access request cannot *directly* (instantaneously) reach its target. Instead, it follows some access path starting from the external space and leading to the internal target space (Fig. 1).

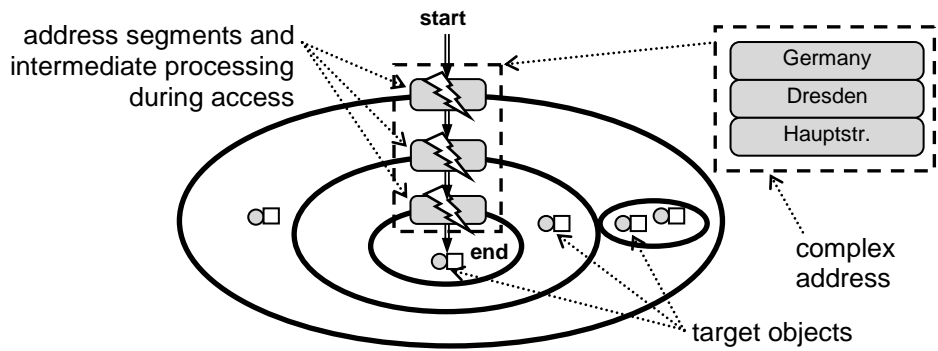

*Figure 1. Program represented as a nested space.*

An object can be accessed only by intersecting several intermediate space borders each represented by one reference segment. Each intermediate border has some behavior associated with its reference which is automatically triggered for all access requests intersecting this border. For example, a bank account is identified by its number *and* by its bank. To access this object, it is necessary to intersect two borders – the bank and the account – each performing some intermediate actions. The object method is the last step in this sequence and it will be executed after all intermediate actions have been completed. Here we again see that a great deal of the program functionality is concentrated on space borders which are described by references while object methods may account for a relatively small portion of the overall complexity. The goal of COP in this sense can be viewed as providing means for describing nested virtual address spaces where objects exist. In other words, we describe a nested space with some custom address space and active borders which is supposed to be used as a container for various classes of objects. The containers may influence the behavior of internal objects by intercepting incoming requests.

Representing objects as existing in a hierarchy is the basis of prototype-based programming (PBP) (Borning, 1986; LaLonde et al., 1986; Lieberman, 1986). The difference is that COP is a class-based approach while PBP does not uses classes. In addition, objects in PBP are still represented by primitive references while COP introduces the notion of complex reference for that purpose. Complex references in COP are active elements of the program which not only represent the position of an object in the hierarchy but also actively participate in processing all requests to internal objects.

## 3. From Classes to Concepts

### 3.1. Concepts and Dual Methods

*Concepts* in COP III are intended to describe *values*. This means that concept instances are passed by-copy and do not have any other indirect representations like address or reference. For example, the following concept describes a bank account:

```
concept Account {
  char[10] accNo;
  Person owner;
}
```

Instances of this concept are values containing 10 characters in the first field and some value in the second field the structure of which is defined by the `Person` concept.

Each method in a concept may have two implementations: an *incoming* method (marked by the modifier 'in') and an *outgoing* method (marked by the modifier 'out'). Such a pair of incoming and outgoing methods with the same signature is referred to as *dual methods*. For example, the logic for getting the current account balance can be implemented in the `getBalance` method which has two versions (incoming and outgoing):

```
concept Account
  char[10] accNo;
  in double getBalance() {...};
  out double getBalance() {...};
}
```

Concept methods are used as usual without any indication whether it is an incoming or outgoing method and the sequence of their invocation depends on the direction of access. Incoming and outgoing methods correspond to reference and object methods, respectively, in the previous version of COP (Savinov, 2007; Savinov, 2008; Savinov, 2009b).

### 3.2. References and Objects

An important assumption in COP is that references are values and hence concept instances (values) can be interpreted as references which indirectly represent objects. In other words, a reference in COP is a value which can provide access to an object. Objects are defined as *functions* of references which return the same output value for the same input reference. Thus an object is identified by its reference and its object fields are defied as functions of this reference. Therefore we say that values are accessed directly while objects are accessed indirectly. If functions are not defined then we get a normal value, that is, values are a particular case of objects without associated functions.

To model objects by means of concepts it is assumed that *object fields are implemented by outgoing methods of concepts* which return the same result for the same reference (concept instance). Syntactically, such methods are defined as setters and getters. For example, bank accounts are uniquely identified by their numbers which are used as references. However, account balance cannot be stored in the reference because it has to be shared among all holders of the reference. Therefore the balance field is defined as an outgoing method which returns balance depending on the account number.

```
concept Account {
  char[10] accNo;
  out double balance {
    get { return func(accNo); }
  }
}
```

Here we effectively defined a new object field, called `balance`, which can be used as usual:

```
Account acc = getAccount("Name");
double currentBalance = acc.balance;
```

If there is no need in having custom references then they can be inherited from some kind of primitive reference provided by the run-time environment as it is done on OOP.

## 4. From Inheritance to Inclusion

### 4.1. Concept Inclusion

Concepts in COP exist in a hierarchy. However, COP uses a new relation, called *inclusion*, for defining new concepts based on already existing concepts. If concept B is included in concept A then A is referred to as a *super-concept* and B is referred to as a *sub-concept*. The main difference of inclusion from inheritance is that inclusion represents containment relation while inheritance is considered a particular case of containment. In other words, if classes in OOP are connected by means of IS-A relation, then concepts in COP are connected via IS-IN relation. For example, since any bank account always exists within some bank, we describe this relationship by means of inclusion (denoted by the keyword 'in'):

```
concept Bank
  char[12] bankCode;
}
concept Account in Bank {
  char[10] accNo;
}
```

Instances of the `Account` concept will extend instances of the `Bank` concept, that is, an instance of `Account` is a value with additional fields attached to an instance of `Bank`.

Inclusion is used to describe hierarchical address spaces similar to postal addresses or computer names. This mechanism is based on the important conceptual assumption that *extended values describe relative references*. This means that a sub-concept describes the structure of addresses relative to its super-concept. And a child instance (extension) is said to exist in the domain (also called context or scope) of its parent instance. In the above example, any variable of the `Account` concept will contain a *complex reference* consisting of two segments: a parent bank reference and a child account reference. Thus concepts and inclusion relation allow us to control what is stored in variables and passed in parameters rather than only the object format in OOP.

### 4.2. Navigating Inclusion Hierarchy

Concept instances in COP exist in a hierarchy which can be traversed in two directions by using two keywords. The conventional '`super`' keyword is used to access the parent element, and a novel '`sub`' keyword is used to access the child element. '`sub`' is analogous to '`inner`' in the Beta programming language (Goldberg et al., 2004). An important distinguishing feature of this navigation is that child instances are accessed using their incoming methods while parent instances are accessed using their outgoing methods. For example, if a method of the `Bank` concept is called from any method of the `Account` concept then an outgoing version of the method will be executed:

```
    concept Account in Bank
      out double getInterest() {
        double rate = super.getInterest();
        return rate + accRate;
      }
    }
```

Here `super.getInterest()` invokes an outgoing method of the `Bank` concept which returns the current interest rate in this bank (the same for all accounts of this bank). An incoming version of this method might return a different interest rate for external calls. Note also that the `getInterest` method of the `Account` concept is an outgoing method and hence it is accessible only from sub-concepts.

A concept can be thought of as a border between the internal domain consisting of its sub-concepts and the external domain consisting of all other concepts. Then any instance intercepts incoming requests by using incoming methods and outgoing requests are intercepted by outgoing methods. This also can be viewed as a visibility rule where outgoing methods are visible from inside and incoming methods are visible from outside. It is analogous to the passport control system at airports where arriving and departing passengers pass through different gates with different procedures.

Object hierarchy is modeled by means of outgoing methods the result of which depends on this and parent instances. In the case of the same parent, outgoing methods of different children will produce different results which are interpreted as different object field values. For example, assume that one bank object has many account objects with the persistent state stored in some database. Account balance could be then defined as follows:

```
    concept Account in Bank {
      char[10] accNo;
      out double balance {
        get {
          Connection db = super.getConn();
          return db.load("balance", accNo);
        }
        set {
          Connection db = super.getConn();
          db.save("balance", accNo, value);
        }
      }
    }
```

Here the `balance` field is defined as an outgoing method (via one setter and one getter). Account balance depends on the current bank which provides connection to the database (so different banks store their data in different databases). The balance also depends on the current account number which is used as a primary key when getting values from the database. Importantly, these are only implementation details but logically all objects exist in a hierarchy where each bank has many accounts. We can read balances and update balances using account references (consisting of several segments). And these operations will be logically correct because their result depends only on the references.

### 4.3. Inheritance

Inheritance is a language mechanism for describing new objects by *reusing* already existing object descriptions. Inheriting concept fields in COP works precisely as in the classical case: child fields are simply attached to the parent concept fields. In this way we can extend values by adding more fields to them. For example, if concept `Point` has two fields `x` and `y` then we can define a new concept `Point3D` which has an additional field `z`:

```
concept Point { int x; int y; }
concept Point3D in Point { int z; }
```

The classical model for object extension can be obtained if the child concept has no fields. Since the reference is empty, only one child instance can exist within one parent (just because children cannot be distinguished). In this case, we can think of child object fields as simply extending the parent fields. For example, if we need to define a bank account with some additional property then it can be done as follows:

```
concept BonusAccount in Account {
  out double bonus; // Object field
}
```

It is equivalent to conventional class and class inheritance. Any instance of this class will get its own parent segment with an additional `bonus` field defined in this concept.

In the general case, inheritance is implemented by reusing parent outgoing methods. More specifically, child outgoing methods are implemented using parent outgoing methods which are called via `super` keyword. This inheritance model is similar to that in prototype-based programming where behavior defined in a parent object (prototype) is shared among and reused by all child objects (Stein, 1987).

### 4.4. Polymorphism

Polymorphism allows an object of a more specific type to be manipulated generically as if it were of a base type. For example, if we declare a variable as having the type `Account` then polymorphism allows us to apply to it the method `getBalance` even though it stores a reference to a more specific type like `BonusAccount`. There exist different approaches to implementing polymorphic behavior but the currently dominating strategy follows the principle that *child object methods have precedence over (override) parent object methods*. In other words, if we define a child method then it will have precedence over the parent methods. If the child still needs some parent functionality then it has to explicitly use it by means of a super call. For example, if the `Button` concept has to provide a more specific implementation of the `draw` method (than its parent `Panel` concept) then it is implemented as follows:

```
concept Panel {
  out void draw() {
    fillBackground();
  }
}
concept Button extends Panel {
  out void draw() {
    super.fillBackground ();
    drawButtonText("MyButton");
  }
}
```

In addition to this classical direct overriding strategy for implementing polymorphism, COP introduces an *inverse overriding strategy*. The main principle of this strategy is that *parent incoming methods have precedence over (override) child incoming methods*. It is useful when it is necessary to control access to the internal scope. Once a parent reference method got control it can decide how to continue. Normally, parent methods perform some actions and then delegate the method call further to its child reference.

For example, the parent class can always fill panel background and then the child method is called in order to add (inject) more specific behavior:

```
concept Panel {
  in void draw() {
    fillBackground();
    sub.draw();
  }
}
concept Button in Panel {
  in void draw() {
    drawButtonText("MyButton");
  }
}
```

Note that here we inject some more specific behavior from within the parent incoming methods instead of injecting more general (parent) behavior from within the child (direct overriding).

The complete sequence of access via dual methods is shown in Fig. 2. It is assumed that concept SavingsAccount is included in concept Account which in turn is included in concept Bank. The left part shows access on incoming methods and the right side shows access on outgoing methods. If a method is applied to such a complex reference then the parent incoming method intercepts it and then the same method of the child element is called. Thus we can move down through incoming methods using keyword 'sub'. At some moment the incoming method can call an outgoing method and the process switches to the right half of the diagram. Here the process propagates upward through the outgoing methods using keyword 'super'.

The inverse overriding strategy is analogous to the idea of treating sub-classes as behavioral extensions to their super-classes in the Beta programming language (Kristensen et al., 1987; Madsen & Møller-Pedersen, 1989) where super-classes provide generic behavior which is extended using the keyword inner rather than overridden. Both strategies describe behavior incrementally by executing some operations and then sending a request for further processing either to the parent or child object. The difference between them is only in the direction of delegation which is also similar to the mechanism of capturing and bubbling in JavaScript. What is new in COP is that these two strategies are combined using the mechanism of dual methods which effectively isolates two directions for method call propagation.

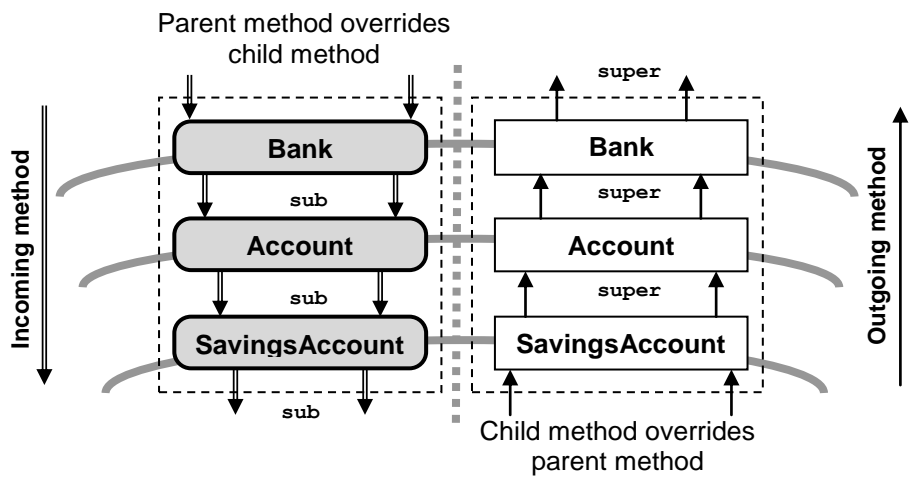

*Figure 2. Two overriding strategies.*

## 4.5. Cross-Cutting Concerns

Complex programs have functions which are scattered throughout the whole source code. Such program logic that spans the whole program is referred to as a cross-cutting concern and is known to produce numerous problems in software development. Aspect-oriented programming (AOP) (Kiczales, 1997) is the most wide spread approach to modularizing cross-cutting concerns which introduces an additional programming construct, called aspect. Aspects are orthogonal to the classes and behavior defined in aspects is injected into points defined in the class hierarchy. In this sense, aspects and classes play different roles; they are not completely unified as well as not completely independent.

COP proposes a novel solution to this problem which is based on the ability of parent methods to intercept access to child methods. Thus cross-cutting concerns are modularized in parent incoming methods and this functionality is injected in child methods. Effectively, this mechanism allows using parent incoming methods as wrappers for child methods so that some functions are guaranteed to be executed for each access while target (child) objects are unaware of this intervention. In terms of spaces, cross-cutting concerns are thought of as functions associated with space borders and automatically triggered for each incoming request passing the border. For example, if we would like to log any access from outside to account balances then this cross-cutting concern is implemented in the `getBalance` incoming method:

```
concept Bank {
  in double getBalance() {
    logger.Debug("Balance accessed.");
    return sub.getBalance();
  }
}
```

Interestingly, the notion of cross-cutting concern can be also applied to outgoing methods which means that one and the same logic is executed for all outgoing requests. For example, if banks have some reserves and they want to log all accesses to this property from inside then it is implemented as an outgoing method:

```
concept Bank {
  protected out double reserves;
  out double getReserves() {
    logger.Debug("Reserves accessed.");
    return this.reserves;
  }
}
```

Now any access to the bank reserves from any child object (like `Account` methods) will be logged. Obviously, this pattern is easily implemented in OOP. We mention it in order to emphasize that cross-cutting behavior has dual nature which is modularized in incoming and outgoing methods.

## 5. Future Research Directions

In future this approach will be developed in the direction of defining concrete programming languages and introducing elements of concept-oriented programming in existing languages. Another direction for research consists in integrating this approach with the concept-oriented data model (COM) (Savinov, 2009a; Savinov, 2011a) which is based on the same principles. The main challenge is to unify programming with data modeling and querying. In particular, the goal is to further develop the concept-oriented query language (Savinov, 2005b; Savinov, 2006; Savinov, 2011b) in the direction of general-purpose programming languages. Identity modeling in COM is analogical to COP and relies on concepts and inclusion hierarchy. Any data item is uniquely identified by its complex reference from a virtual address space which can be stored in other data items as a property. COM uses the formalism of nested ordered sets for describing data semantics and such operations as projection and de-projection for data manipulations.

## 6. Conclusion

Concept-oriented programming introduces a new programming construct, concept, and a new relation, inclusion, which generalize classes and inheritance, respectively. These new constructs lead to revisiting some basic notions and mechanisms like class, inheritance, referencing, polymorphism, and cross-cutting concerns. Now the world is still described using a hierarchy but the concept-oriented approach reflects the dual nature of things which consist of one entity and one identity. Such a generalization is informally analogous to introducing complex numbers in mathematics which also have two constituents: a real part and an imaginary part. These two constituents are always manipulated as one whole and this makes mathematical expressions much simpler and more natural. The same effect we get in programming when concepts are introduced: program code gets simpler and more natural because two sides or flavors (references and objects) are manipulated as one whole.

The main advantage of COP is in the ability to describe many existing programming patterns and techniques by using only a few basic notions and generalized constructs. In particular, concept-oriented programming can be used to solve the following general problems:

- Access control. Objects in COP are accessed via their references and references encapsulate the logic of access. This mechanism has numerous applications where the logic of access has to be described separately from the normal business logic. For example, it can be used for security checks, transaction control and persistence management. The benefit is that this logic can be encapsulated in references instead of being integrated with business functions of objects.

- Data typing. COP eliminates the need in having two separate constructs for modeling values and objects like struct and class. COP combines by-value and by-reference semantics by eliminating the need in having special modifiers like by-ref and by-val in some programming languages for specifying the way objects are passed. Instead, this separation is specified at the level of data types.

- Object hierarchies. COP combines object-oriented programming with prototype-based programming by making it possible to use class-based approach for modeling object hierarchies. Note that incoming and outgoing methods in COP correspond to two directions for interaction propagation along object hierarchy, called capturing and bubbling in JavaScript.

- Cross-cutting concerns. COP can be viewed as an alternative approach to modularizing cross-cutting concerns which uses concepts (generalized classes) for two purposes: describing conventional object methods and describing intermediate behavior. COP is simpler and more natural than AOP because the injection of intermediate actions is performed along the object hierarchy rather than in arbitrary points of the program.

COP is a step towards developing a unified approach to programming which supports such existing paradigms as object-oriented programming, aspect-oriented programming, prototype-based programming, functional programming. Taking into account the simplicity and generality of COP, it seems rather perspective direction for further research and development in the area of programming paradigms.

Dijkstra, E.W. (1976). *A Discipline of Programming*. Prentice Hall.

Goldberg, D.S., Findler, R.B., Flatt, M. (2004). Super and inner: together at last! *Proc. OOPSLA'04* (pp. 116-129).

Kiczales, G., Rivieres, J., Bobrow, D.G. (1991). *The Art of the Metaobject Protocol*. MIT Press.

Kiczales, G., Ashley, J.M., Rodriguez, L., Vahdat, A., Bobrow, D.G. (1993). Metaobject protocols: Why we want them and what else they can do. In Paepcke, A. (ed.) *Object-oriented programming: The CLOS perspective* (pp. 101-118), MIT Press.

Kiczales, G., Lamping, J., Mendhekar, A., Maeda, C., Lopes, C., Loingtier, J.-M. and Irwin, J. (1997). Aspect-oriented programming. *Proc. ECOOP'97* (pp. 220-242).

Kristensen, B.B., Madsen, O.L., Moller-Pedersen, B., Nygaard, K. (1987). The Beta programming language. In *Research Directions in Object-Oriented Programming* (pp. 7-48), MIT Press.

Lieberman, H. (1986). Using prototypical objects to implement shared behavior in object-oriented systems. *Proc. OOPSLA'86* (pp. 214-223).

LaLonde, W.R., Thomas, D.A., Pugh, J.R. (1986). An exemplar based Smalltalk. *Proc. OOPSLA'86* (pp. 322-330).

Madsen, O.L., Møller-Pedersen, B. (1989). Virtual classes: A powerful mechanism in object-oriented programming. *Proc. OOPSLA'89* (pp. 397-406).

Monson-Haefel, R. (2006). *Enterprise JavaBeans*, O'Reilly.

Savinov, A. (2005a). Concept as a generalization of class and principles of the concept-oriented Programming. *Computer Science Journal of Moldova (CSJM)*, **13**(3), 292-335.

Savinov, A. (2005b). Logical navigation in the concept-oriented data model. *Journal of Conceptual Modeling, 36*. Retrieved December 5, 2012, from http://conceptoriented.org/savinov/publicat/jcm_05.pdf

Savinov, A. (2006). Query by Constraint Propagation in the Concept-Oriented Data Model. *Computer Science Journal of Moldova*, **14**(2), 219-238.

Savinov, A. (2007). *An approach to programming based on concepts*. Institute of Mathematics and Computer Science, Academy of Sciences of Moldova, Technical Report RT0005, 49 pp.

Savinov, A. (2008). Concepts and concept-oriented programming. *Journal of Object Technology (JOT), **7***(3), 91-106.

Savinov, A. (2009a). Concept-oriented model. In V.E. Ferraggine, J.H. Doorn, & L.C. Rivero (Eds.), *Handbook of research on innovations in database technologies and applications: Current and future trends* (2nd ed., pp. 171-180). IGI Global.

Savinov, A. (2009b). Concept-Oriented Programming. *Encyclopedia of Information Science and Technology, 2nd Edition*, Editor: Mehdi Khosrow-Pour, IGI Global, 672-680.

Savinov, A. (2011a). Concept-oriented model: Extending objects with identity, hierarchies and semantics. *Computer Science Journal of Moldova (CSJM)*, **19**(3), 254-287.

Savinov, A. (2011b). Concept-Oriented Query Language for Data Modeling and Analysis, In L. Yan and Z. Ma (Eds.), *Advanced Database Query Systems: Techniques, Applications and Technologies*, IGI Global, 85-101.

Savinov, A. (2012a). Concept-oriented programming: Classes and inheritance revisited. *Proc. 7th International Conference on Software Paradigm Trends (ICSOFT 2012)* (pp. 381-387).

Smaragdakis, Y., Batory, D. (1998). Implementing layered designs with mixin-layers. *Proc. ECOOP'98* (pp. 550-570).

Stein, L.A. (1987). Delegation is inheritance. *Proc. OOPSLA'87* (pp. 138-146).
11